\documentclass[%
 reprint,
 amsmath,amssymb,
 aps,
]{revtex4-1}

\usepackage{graphicx}
\usepackage{dcolumn}
\usepackage{bm}

\usepackage{diagbox}
\usepackage{multirow, array}
\usepackage[caption=false]{subfig}
\usepackage{physics, slashed}
\usepackage{bbm}
\usepackage{hyperref}

\newcommand{\crea}[1]{\hat{#1}^{\dagger}}
\newcommand{\mum}[0]{\mu\text{m}}

\hypersetup{colorlinks=true, citecolor=blue}

\begin{document}


\title{Witnessing the non-classical nature of gravity in the presence of unknown interactions}

\author{Hadrien Chevalier}\thanks{hadrien.chevalier17@imperial.ac.uk}
\author{A. J. Paige}\thanks{a.paige16@imperial.ac.uk}
\author{M. S. Kim}\thanks{m.kim@imperial.ac.uk}
\affiliation{%
QOLS, Blackett Laboratory, Imperial College London, SW7 2AZ, United Kingdom
}%

\date{\today}

\begin{abstract}
General relativity as a classical field theory does not predict gravitationally induced entanglement, as such, recent proposals seek an empirical demonstration of this feature which would represent a significant milestone for physics. We introduce improvements to a spin witness protocol that reduce the highly challenging experimental requirements. After rigorously assessing approximations from the original proposal [S. Bose et al. Phys. Rev. Lett. 119, 240401 (2017)], we focus on entanglement witnessing. We propose a new witness which greatly reduces the required interaction time, thereby making the experiment feasible for higher decoherence rates, and we show how statistical analysis can separate the gravitational contribution from other possibly dominant and ill-known interactions. We point out a potential loophole and show how it can be closed using state tomography.
\end{abstract}

\pacs{Valid PACS appear here}
\maketitle


\section{\label{sec:level1}Introduction}

Gravity, the earliest of the four known fundamental interactions to be studied, is currently best described by Einstein's general relativity (GR), a classical field theory~\cite{wald2010general}. Although GR has proven to be extremely robust to experimental tests, to the point that there is still no convincing empirical evidence going against its predictions~\cite{will2006confrontation}, there are many reasons why the coexistence of GR and quantum theory leaves problematic open questions~\cite{hossenfelder2013possibility, tilloy2019does}. Hence finding evidence for non-classical gravity would constitute a major breakthrough in modern theoretical physics.

The pursuit of post-GR evidence has given rise to novel ideas and experimental proposals ~\cite{pikovski2012probing, stickler2018probing, howl2019exploring, krisnanda2020observable, miao2019quantum, carlesso2019testing, altamirano2018gravity, howl2020testing}. In particular recent works suggesting tests for gravitationally generated entanglement~\cite{marletto2017gravitationally, bose2017spin} have drawn substantial attention~\cite{howl2020testing, khosla2018classical, belenchia2018quantum, hall2018two, christodoulou2018possibility, marletto2019answers, carney2019tabletop, christodoulou2019possibility}. In this paper we focus on the spin witness approach version~\cite{bose2017spin}, which we shall refer to as the spin witness protocol (SWP). The proposal features microdiamonds with embedded NV centers in two Mach-Zehnder setups that are positioned close to one another, inducing phase shifts between ideal couples of position eigenstates, which result in entangled spin density matrices. It is based on the quantum information theoretic fact that entanglement cannot be increased through a local operations and classical communication~\cite{horodecki2009quantum}. 

Although the claim about witnessing the \emph{quantum} nature of gravity as presented originally has been subject to some debate~\cite{hall2018two}, it was pointed out that what is tested is a non-classicality of the gravitational field, in the sense that a successful test rules out any framework in which its state is described by (possibly probabilistic) unique (tensorial, vectorial, scalar, etc.) values at each space-time point~\cite{marshman2019locality, marletto2018can}. Regardless of terminology, the realization of such a proposal would be significant, as far as gravitationally induced entanglement would demonstrate a behaviour (superposition of space-time geometry) that is not predicted by GR~\cite{carlesso2019testing, christodoulou2018possibility, christodoulou2019possibility}.

From an experimental point of view there are several difficulties in the splitting and refocusing operations, such as control pulse timing, particle rotation~\cite{stickler2016spatio}, radiation and spin decoherence~\cite{bateman2014near}, along with diamagnetic properties of diamond, overheating and loading issues~\cite{pedernales2019motional, frangeskou2018pure, bykov2019direct}. These issues are important but they are not the focus of this work. Instead, we address theoretical questions that underpin the protocol independent of our ability to overcome the immense experimental challenges. This work makes any such experiment more viable.

We critically revisit the SWP, by introducing new theoretical considerations and closing a potential loophole. We begin in Sec.~\ref{sec:rigorousSWP} by showing that the original treatment~\cite{bose2017spin} with position eigenstates is a valid approximation for all intents and purposes: having realistic coherent or thermal states results in negligible corrections. More importantly, we go on to demonstrate in Sec.~\ref{sec:witnesses} that by introducing a better entanglement witness, entanglement can be revealed in a shorter free fall time, which improves the experiment's tolerance to decoherence. Since empirical results are statistical in nature, we adopt a likelihood ratio approach in Sec.~\ref{sec:hypotheses} which allows, from repeated witness measurements, to draw conclusions on the non-classicality of gravity in regimes dominated by non-gravitational forces, such as Casimir-Polder (CP) interactions. The method works even without having exact knowledge of their coupling strength, as shown in Sec.~\ref{sec:unknownforces}.  Finally, in Sec.~\ref{sec:reconstruction} we point out that although witness experiments can provide convincing evidence, a fully rigorous certification of gravitationally driven entanglement would require the knowledge of an entanglement monotone. We illustrate a solution to this by simulating state tomography, and we give figures for the number of repetitions required in different settings.


\section{Arbitrary product states in the spin witness protocol}\label{sec:rigorousSWP}

\begin{figure}
    \centering
    \includegraphics[scale=0.55]{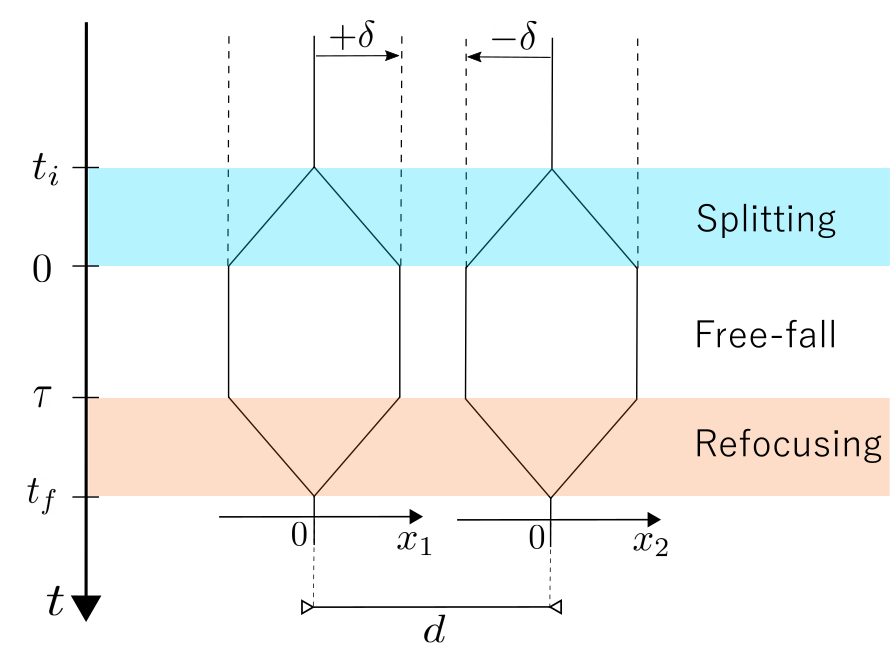}
    \caption{Illustration of the SWP protocol. The two systems have two position coordinates the origins of which are separated by a distance $d$.}
    \label{fig:bosediagram}
\end{figure}

The SWP setup~\cite{bose2017spin} is illustrated in Fig.~\ref{fig:bosediagram}, and consists of two Mach-Zehnder interferometers. The system is initially in state $\rho^i$ which is a product of two motional states in a superposition of spin states, separated by distance $d$. At the end of the splitting, which is a spin-controlled spatial displacement by $\pm \delta$, the system is in state $\rho(0)$ which is two spatial superpositions. The free-fall duration is labelled by $\tau$, after which the state is $\rho(\tau)$. The non-adaptive refocusing merges the positional superpositions without taking into account drift during free-fall, and results in a final state that depends on the free-fall duration $\rho^f(\tau)$.

The original proposal implicitly adopts two theoretical simplifications: using position eigenstates, and assuming that tracing out the motional degree of freedom after refocusing does not completely decohere the spin state. Although useful for illustrating the key ideas, it should be noted that these are contradictory. Should the states be rigorously delta-distributed in position-space, then drift can absolutely not be neglected. For the sake of completeness, we derive results presented in~\cite{bose2017spin} without using position eigenstates, but instead start from products of two arbitrary trapped motional states. We find the resulting corrections to the position eigenstate approximation (PEA) to be negligible for a reasonable range of trapping frequencies and temperatures.

As in the original proposal~\cite{bose2017spin}, we assume that the splitting and refocusing operations can be done in a short time compared to the free-fall duration, and use a Newtonian potential for gravity. Since $d/c \ll \tau$, this is a completely valid~\cite{belenchia2018quantum} static limit to the fully general relativistic description, which formulates the same predictions in the case of superposition of geometries~\cite{christodoulou2019possibility}. It should be stressed that the conclusion on non-classicality through entanglement growth, is model agnostic~\cite{marletto2019answers, marletto2020witnessing}.

Consider two identical particles of mass $m$ that are initially in a product of two arbitrary motional states and in a superposition of spin states $\rho^i = (\pi_1\otimes \rho_+)\otimes(\pi_2\otimes \rho_+),$ where $\rho_{+}=\frac{1}{2}(\ket{s_L}+\ket{s_R})(\bra{s_L}+\bra{s_R}),$ and this spin notation will provide labels for the displacements. As in~\cite{bose2017spin}, it is assumed the spin-controlled spatial splitting can be performed such that that there is no mean momentum at the beginning of the free fall. The splitting operation reads $\left( \hat{D}_L\otimes \dyad{s_L}{s_L} + \hat{D}_R\otimes \dyad{s_R}{s_R}\right)^{\otimes 2}$, where $\hat{D}_{\mu}=\hat{D}(\kappa_{\mu}) = e^{\kappa_{\mu}\crea{a} - \kappa_{\mu}^*\hat{a}}$ is the displacement operator, and $\kappa_R = -\kappa_L = \kappa \in \mathbb{R}_+$. We denote $\delta$ the physical distance by which the state is displaced, so $\kappa = \delta\sqrt{m\omega/2\hbar}$ where $\omega$ is some initial trap frequency.

In the noiseless case, the final state obtained at the end of the free-fall after the refocusing operation, which is the Hermitian conjugate of the splitting, reads
\begin{equation}\begin{aligned}
    \rho^f(\tau) = \frac{1}{4}\sum\limits_{\alpha\beta\mu\nu} (\crea{D}_{\alpha}\otimes&\crea{D}_{\beta})\hat{U}_{d}(\hat{D}_{\alpha}\otimes\hat{D}_{\beta}) \rho^i\\& \times(\hat{D}_{\mu}\otimes\hat{D}_{\nu})\hat{U}_{d}(\crea{D}_{\mu}\otimes\crea{D}_{\nu}),
    \end{aligned}
\end{equation}
where the sum is performed over $(\alpha, \beta, \mu,\nu)\in\{L,R\}$, $\hat{U}_{d}$ is the propagator generated by the Hamiltonian $\hat{H}_d = (\hat{p}_1^2+\hat{p}_2^2)/2m - Gm^2/(d+\hat{x}_2 - \hat{x}_1)$ for time $\tau$, and $G$ is the gravitational constant. The displacement operations amount to a shift of origins for the position operators, which can be absorbed in the separating distance $d$. Explicitly, $(\crea{D}_{\mu}\otimes\crea{D}_{\nu})\hat{U}_{d}(\hat{D}_{\mu}\otimes\hat{D}_{\nu})    = \hat{U}_{d_{\mu\nu}}$ where $d_{\mu\nu} = d - \delta_{\mu} + \delta_{\nu} \in \{d - 2\delta, d, d+2\delta\}$. Thus, instead of having three distinct relative displacements, we deal with three distinct separations and propagators acting on the same initial state. Then, by pulling out the order zero potential term $-Gm^2/d_{\mu\nu}$ from the Hamiltonian, the matrix elements $s_{\alpha\beta\mu\nu}$ of the reduced spin state $\Tr_{\text{motion}}[\rho^f(\tau)] = \sum_{\alpha\beta\mu\nu}s_{\alpha\beta\mu\nu}\dyad{s_{\alpha} s_{\beta}}{s_{\mu} s_{\nu}}$ reduce to
\begin{equation}\label{eq:SDMelements}
    s_{\alpha\beta\mu\nu} = \frac{1}{4} \exp[\frac{-iGm^2\tau}{\hbar}Q^{(1)}_{\alpha\beta\mu\nu}] \Tr[{\hat{\slashed{U}}_{\alpha\beta}}^{\dagger}\hat{\slashed{U}}_{\mu\nu}(\pi_1\otimes \pi_2)],
\end{equation}
where $\forall n\in\mathbb{N}, \  Q^{(n)}_{\alpha\beta\mu\nu} = \frac{1}{d_{\alpha\beta}^n} - \frac{1}{d_{\mu\nu}^n}$, and where $\hat{\slashed{U}}_{\alpha\beta}$ are propagators generated by the Hamiltonians with no order zero potentials. Essentially, the propagation of the full system is equivalent, up to a shift of position operator origins, to a sum of four pairwise evolutions, three of which are distinct. The PEA result adopted in~\cite{bose2017spin} can be arrived at by discarding the remaining trace.

Up to now, our approach is valid for position coordinates in the open disc of convergence of the analytic expansion of the potential term. We now restrict ourselves to the case where $|{x}_2-{x}_1| \ll d - 2\delta$ and inspect some results we might infer with a truncated potential. With an order $1$ truncation the Baker-Campbell-Hausdorff identity gives
\begin{equation}\label{eq:propproduct}
\begin{aligned}
    \hat{\slashed{U}}_{\alpha\beta}^{\dagger}\hat{\slashed{U}}_{\mu\nu} = e^{-i\tau^3G^2m^3Q^{(4)}_{\alpha\beta\mu\nu}/6\hbar}\hat{D}\left(\theta_{\alpha\beta\mu\nu}\right)\otimes \hat{D}\left(-\theta_{\alpha\beta\mu\nu}\right), 
\end{aligned}
\end{equation}
where $\theta_{\alpha\beta\mu\nu} = \frac{GmQ^{(2)}_{\alpha\beta\mu\nu}\tau}{\sqrt{2}}\left[\frac{\tau}{2}\sqrt{\frac{m\omega}{\hbar}} - i\sqrt{\frac{m}{\hbar\omega}}\right]$. As expected from classical mechanics, the two particles are displaced towards one another and acquire opposite momenta.

For two initially identical motional states $\pi_1 = \pi_2 = \pi$,
\begin{equation}\label{eq:thermal1}\begin{aligned}
    \Tr[{\hat{\slashed{U}}_{\alpha\beta}}^{\dagger}\hat{\slashed{U}}_{\mu\nu}\pi^{\otimes 2}] =& e^{-i\tau^3G^2m^3Q^{(4)}_{\alpha\beta\mu\nu}/6\hbar}\\
    & \times e^{-|\theta_{\alpha\beta\mu\nu}|^2}\left(C_N\left(\frac{\theta_{\alpha\beta\mu\nu}}{2}\right)\right)^2,
\end{aligned}
\end{equation}
where $C_N:\lambda\longmapsto \Tr[\pi e^{\lambda\crea{a}}e^{-\lambda^*\hat{a}}]$ is the normally ordered characteristic function of $\pi$~\cite{gerry2005introductory}. From this, the spin density matrix elements for an initial thermal state with $\langle\hat{N}\rangle = \overline{n}$ is deduced to obey
\begin{equation}\label{eq:thermal2}\begin{aligned}
    s_{\alpha\beta\mu\nu} = \frac{1}{4}&e^{-iGm^2\tau Q^{(1)}_{\alpha\beta\mu\nu}/\hbar}\\ &\times e^{-iG^2m^3\tau^3Q^{(4)}_{\alpha\beta\mu\nu}/6\hbar}e^{-(\frac{\overline{n}}{2} + 1)|\theta_{\alpha\beta\mu\nu}|^2}.
\end{aligned}
\end{equation}
 The first phase factor is the only term considered in the original SWP. The second phase factor is a first order phase correction. The third factor is a first order decoherence effect due to drift. Proofs of Eqs.~\eqref{eq:thermal1} and~\eqref{eq:thermal2} can be found in appendix~\ref{sec:proofs1}. 

To facilitate comparison, we work with the parameters of the original proposal~\cite{bose2017spin}, $(d = 400 \ \mum, \ \delta = 125 \ \mum, \ m  = 10^{-14} \ \text{kg})$ and with a sensible trapping frequency $\omega = 10^3 \ \text{Hz}$~\cite{hsu2016cooling}. The fastest oscillating terms ($\alpha\beta\mu\nu = LRRL$) have $|Q^{(1)}| \sim 3.6\times 10^3 \ \text{m}^{-1}$ and $|Q^{(4)}| \sim 6.2\times 10^{14} \ \text{m}^{-4}$. The first phase reaches unit radian after a characteristic free-fall duration $\tau \sim 4.4 \ \text{s}$. After $10$ seconds of free-fall, the phase correction is approximately $4\times 10^{-12} \ \text{rad}$ and the decoherence factors are $\exp(-2.8\times 10^{-8})$ and $\exp(-0.014)$ respectively for zero temperature and $T = 7.6 \ \text{mK} \ (\overline{n} = 10^6)$. This shows the PEA to be valid for all intents and purposes.

In the PEA, the reduced spin state can be considered to be pure and reads, up to a global phase, $\ket{\psi_s(\tau)} = \frac{1}{2}(\ket{00} + e^{i\Delta\phi_{LR}}\ket{01} + e^{i\Delta\phi_{RL}}\ket{10} + \ket{11})$ where $\Delta\phi_{\mu\nu} = Gm^2\tau(\frac{1}{d} - \frac{1}{d_{\mu\nu}})$. This is the form that was directly posited in the original proposal~\cite{bose2017spin}, and we shall take it as our noiseless state.

It was furthermore argued that the CP interactions could be neglected with the original parameters. One can indeed write down the potential from~\cite{casimir1948influence} as $\hat{V}^C_{\mu\nu} = - \alpha(R,\epsilon)/(d_{\mu\nu}+\hat{x}_2-\hat{x}_1)^7$, where $ \alpha(R,\epsilon) = \left(\frac{\epsilon - 1}{\epsilon + 2}\right)^2\frac{23 \hbar c R^6}{4\pi}$ depends on the radius $R$ of the microspheres and their relative permittivity $\epsilon$. In this case the spin density matrix elements read
\begin{equation}\label{eq:altSDM}\begin{aligned}
    s_{\alpha\beta\mu\nu} =\frac{1}{4}& \exp(\frac{-it}{\hbar}\left(Gm^2 Q^{(1)}_{\alpha\beta\mu\nu} + \alpha Q^{(7)}_{\alpha\beta\mu\nu}\right))\\ &\times \Tr[{\hat{\slashed{U}}_{\alpha\beta}}^{\dagger}\hat{\slashed{U}}_{\mu\nu}(\pi_1\otimes \pi_2)].
\end{aligned}
\end{equation}
With $R \approx 10^{-4} \ \text{m}$, which roughly corresponds to a diamond microsphere of mass $10^{-14} \  \text{kg}$ and $\epsilon \approx 5.7$, the most rapidly evolving terms have a gravity frequency of $0.226 \ \text{Hz}$ and a CP frequency $0.016 \ \text{Hz}$. As claimed originally, one may argue that the CP interaction is negligible compared to the gravitational coupling if the closest approach is kept above roughly $200 \ \mum$.

Since we will show how to overcome the CP closest approach limit given in~\cite{bose2017spin}, it is worth mentioning that the PEA is still valid in a smaller setup separation. If we decrease the separation distance $d$ from $450 \ \mum$ to $350 \ \mum$, the closest approach is then $100 \ \mum$  such that for the $\ket{RL}$ pair the regime is dominated by CP coupling, which becomes roughly $4$ times as strong as the gravitational coupling. The characteristic free-fall duration is lowered to $\tau \approx 54 \ \text{ms}$. After $1$ second of free-fall, the phase correction is approximately $7.03\times 10^{-14} \ \text{rad}$ and the decoherence factors are $\exp(-5\times 10^{-11})$ and $\exp(-2.5\times 10^{-6})$ respectively for zero temperature and $T = 7.6 \ \text{mK} \ (\overline{n} = 10^6)$. Those corrections to the PEA are indeed still negligible.

\section{Witnesses to verify gravitationally induced entangled states}\label{sec:witnesses}
We now introduce the non-separability condition from~\cite{bose2017spin}, and give its corresponding entanglement witness. We propose a new entanglement witness that is optimal in the sense that it theoretically can detect entanglement for arbitrarily short free-fall durations $\tau$.

\begin{figure}
    \centering
    \includegraphics[scale=0.55]{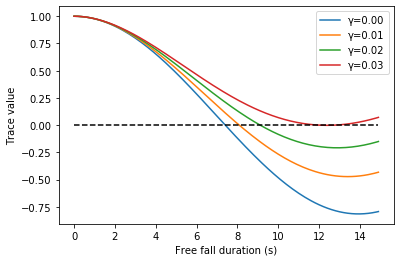}
    \caption{$\Tr(W_0\dyad{\psi_s}{\psi_s})$, where $W_0$ is the entanglement witness put forward in~\cite{bose2017spin}, as a function of free-fall time with decoherence rates $\gamma\in\{0, 0.01, 0.02, 0.03\}$. At $\gamma = 0.03$ the witness can no longer reveal entanglement.}
    \label{fig:originalwitness}
\end{figure}

From the pure spin state resulting from the PEA, one can read off $\Delta\phi_{LR}+\Delta\phi_{RL} \in \{2n\pi | n\in\mathbb{Z}\}$ as a necessary and sufficient condition for separability. In the original proposal~\cite{bose2017spin}, the condition $|\ev{X\otimes Z} + \ev{Y\otimes Y}| > 1$ is put forward as certifying entanglement, where $X,Y,Z$ denote Pauli operators. This formally corresponds to selecting
\begin{equation}
    W_0 = I\otimes I + X\otimes Z + Y\otimes Y,
\end{equation}
as an entanglement witness~\cite{horodecki2001separability}, as for any separable two qubit state $\rho$, $\Tr(W_0\rho) \geq 0$. In the noiseless case, entanglement is revealed after roughly $8$ seconds of free-fall, as shown in Fig.~\ref{fig:originalwitness}.

\begin{figure}
    \centering
    \includegraphics[scale=0.55]{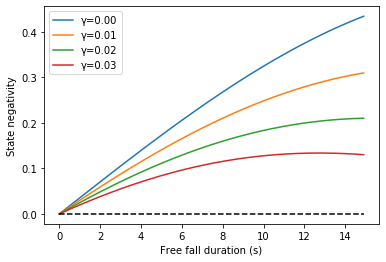}
    \caption{Negativity of the spin state with respect to free-fall time with decoherence rates $\gamma\in\{0, 0.01, 0.02, 0.03\}$. Negativity is an entanglement monotone, and the state is entangled when the negativity is positive.}
    \label{fig:neg}
\end{figure}

Although the order of magnitude for the required free-fall time is promising, it would still correspond to a falling distance of a few $10^2$ meters on Earth, and is still $3$ orders of magnitude above the coherence times observed in cutting edge matter-wave interferometry with much less massive particles~\cite{fein2019quantum}. Adapting the protocol to work with shorter free-fall times makes it more feasible and more robust to decoherence. To illustrate the effect of decoherence, we choose a scattering term that induces an exponential dephasing of local motional states~\cite{kiefer1999decoherence}. We shall denote the off-diagonal damping rate $\gamma$. Explicitly in the local position eigenstate basis $\{\ket{L},\ket{R}\}$, the decoherence after duration $\tau$ acts as a dephasing channel $\pi\longmapsto (1-p)\pi + pZ\pi Z$ where $p = (1 - e^{-\gamma \tau})/2$. The original witness $W_0$ fails to detect any entanglement for $\gamma \geq 0.03 \ s^{-1}$.
The failure of this witness is due to the fact that even in an ideal zero-temperature noiseless scenario, it requires over $8$ seconds of free-fall time for revelation when in fact state negativity~\cite{lee2000partial} $\mathcal{N}(\rho_s) = \sum_{\lambda\in \text{Sp}(\rho_s)\cap \mathbb{R}_-} |\lambda|$ is achieved immediately, as shown in Fig.~\ref{fig:neg}.

To shorten the required interaction time, we build another entanglement witness with a few local Pauli measurements in the spirit of~\cite{riccardi2019optimal}, using the PPT-criterion~\cite{horodecki1997separability}. We assume $\phi = \Delta\phi_{LR} \gg \Delta\phi_{RL}, \Delta\phi$, which amounts to neglecting all but the phase induced by the strongest interacting couple of states. In the PEA, the final spin state reads $\ket{\psi_s(\phi)} = \frac{1}{2}(\ket{00}+\ket{01}+e^{i\phi}\ket{10}+\ket{11})$ or equivalently $\rho_s(\phi) = \dyad{\psi_s(\phi)}{\psi_s(\phi)}$. The eigenstate associated with the negative eigenvalue of the partially transposed $\rho_s^{\Gamma_2}(\phi)$ is $\ket{\chi_{-}(\phi)} = \frac{1}{2}(\ket{00} + ie^{-i\phi/2}\ket{01} - ie^{i\phi/2}\ket{10} - e^{-i\phi}\ket{11})$. At $\phi=0$ one has $4\dyad{\chi_-}{\chi_-}= I\otimes I - X\otimes X + Z\otimes Y - Y\otimes Z$, therefore a witness can be defined as
\begin{equation}
    W_1 = 4\dyad{\chi_-}{\chi_-}^{\Gamma_2} = I\otimes I - X\otimes X - Z\otimes Y - Y\otimes Z.
\end{equation}

\begin{figure}
    \centering
    \includegraphics[scale=0.55]{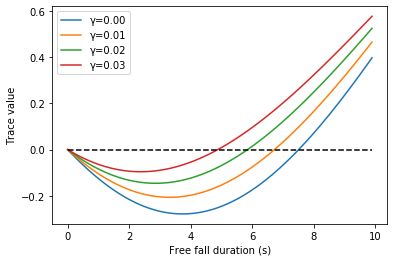}
    \caption{$\Tr(W_1\dyad{\psi_s}{\psi_s})$ as a function of free-fall time with decoherence rates $\gamma\in\{0, 0.01, 0.02, 0.03\}$. The witness in theory can reveal entanglement even for relatively strong decoherence rates if the free-fall time is kept short. This comes at the expense of expectation values that are closer to zero.}
    \label{fig:new_witness}
\end{figure}

This witness reveals entanglement immediately after the start of the free-fall, as shown in Fig.~\ref{fig:new_witness}, as long as the decoherence rate $\gamma$ satisfies $\gamma < (\omega_{RL} + \omega_{LR})/2$ where the $\omega$ are respective coupling strengths $\omega_{\mu\nu}t = \Delta\phi_{\mu\nu}$, as demonstrated in \ref{sec:proofs2}. With the original parameter settings, the witness works in principle for $\gamma < 0.0627 \ \text{s}^{-1}$. Numerics show that the state is in fact not entangled for any higher decoherence rates, hence our witness is in this sense optimal.

Having a witness that can detect entanglement in theory with arbitrarily minute phase accumulation is advantageous, as it decreases the required free-fall duration. However, it raises the question of residual interactions, such as CP coupling. Even if gravity were to be the dominant interaction, small CP couplings would still induce entanglement that will be detected by the new witness. Empirical results are statistical statements, and the presence of additional interactions also should afflict experimental data on the original witness $W_0$. If measurements are performed after a free-fall duration $\tau$ when $\ev{W_0}$ is barely negative, then one must be able to make statistical statements on the impact of non-gravitational interactions on the observed entanglement. The consequences of such an experiment being realized are important enough to require a more rigorous approach to the analysis of empirical results. The effect of a negligible but still existing CP coupling on the entanglement witness and more precisely on the resulting statistics, deserve closer inspection.

\section{Ruling out the absence of gravitational coupling from measurement data}\label{sec:hypotheses}

In order to affirm that the final spin state was induced by a gravitational propagator in the presence of other interactions, such as CP, that we for now assume we have good knowledge of, we adapt the approach developed in~\cite{blume2010entanglement} for entanglement verification. In using such statistical methods, we assume the experiment can be repeated, for instance with particle recycling as outlined in~\cite{stickler2018probing}. The likelihood ratio test being the most powerful test for a given confidence level, according to the Neyman-Pearson lemma~\cite{Neyman1933}, we look at likelihood ratios between
\begin{itemize}
    \item The null hypothesis $H_0$ is: ``\textit{The observed state is entangled state and results from CP interactions without gravity.}"
    \item The alternative hypothesis $H_a$ is: ``\textit{The observed state is entangled and results not only from CP coupling but also from a gravitational interaction.}"
\end{itemize}
In the SWP as originally presented~\cite{bose2017spin}, neglecting CP interactions and using a sub-optimal witness essentially simplifies the hypotheses as entanglement can only be witnessed if $H_a$ is true.

It should be stressed that ruling out $H_0$ in favour of $H_a$ is essentially making two statements. The first one is that the observed empirical data is highly unlikely to have occurred without gravitational coupling. The second, is that the data corresponds to an entangled state. We begin by focusing on ruling out the absence of gravitational contribution.

To obtain the likelihood ratios, in general one can choose to measure a list of bipartite Pauli observables $\underline{\sigma} = [\sigma_1, ..., \sigma_l]$, $N\in\mathbb{N}^*$ times each. Each bipartite observable has $4$ eigenstates. The full list of eigenstates is $\underline{e} = [\ket{e_{11}}, ..., \ket{e_{14}}, \ket{e_{21}}, ..., \ket{e_{l4}}]$.
This defines a $4l$-dimensional probability vector
\begin{equation}
    \underline{p} = [p_{ij}]_{1\leq i \leq l, 1\leq j\leq 4} = [\Tr(\rho \dyad{e_{ij}}{e_{ij}})].
\end{equation}
The data $\underline{n} = [n_{11}, ..., n_{14}, n_{21}, ..., n_{l4}]$ is a list of number of occurrences of measurement outcomes, each corresponding to an obtained eigenstate. The probability of of obtaining the empirical data vector $\underline{n}$ from state $\rho$ is the joint probability distribution
\begin{equation}
    \mathbb{P}(\underline{n}|\rho) = \prod\limits_{ij} p_{ij}^{n_{ij}} \overset{\text{def.}}{=} \mathcal{L}(\rho|\underline{n}),
\end{equation}
and defines the likelihood $\mathcal{L}(\rho|\underline{n})$ of the state $\rho$ given the empirical data vector $\underline{n}$.

The likelihood ratio, assuming the alternative state is obtained at the end of the experiment, reads
\begin{equation}
    \Lambda_a = \frac{\mathcal{L}(\rho_0(\gamma,t)|\underline{n}_a)}{\mathcal{L}(\rho_a(\gamma,t)|\underline{n}_a)},
\end{equation}
where $\rho_0$ is the spin state obtained at the end of the SWP with an exclusively CP induced evolution (null hypothesis state), $\rho_a$ is the spin state obtained under the full CP and gravitational propagator (alternative hypothesis state), and $\underline{n}_a$ is an empirical vector obtained from measurements on $\rho_a$. The free-fall duration $\tau$ is chosen to correspond to the duration after which the witness expectation value is minimal in the alternative hypothesis, so as to maximize the probability of certifying entanglement. This makes $\tau$ dependent on $\gamma$.

We shall use the logarithmic likelihood ratio $\lambda_a = -2\log(\Lambda_a)$, which in the alternative hypothesis reduces to the scalar product $\lambda_a = 2\underline{n}_a.\left(\log(\underline{p}_a) - \log(\underline{p}_0)\right)$, where the subscripts for the probability vectors correspond to the state ($\rho_0$ or $\rho_a$) from which they are constructed.
For the witness measurement, $l = 3$ and the data $\underline{n}_a$ is a vector encoding $3N$ empirical measurement outcomes ($N$ for each of the bipartite Pauli observable $X\otimes X, Y\otimes Z, Z\otimes Y$) on $\rho_a$. High values of $\lambda_a$ strongly support the alternative hypothesis.

\begin{figure}
    \centering
    \includegraphics[scale=0.55]{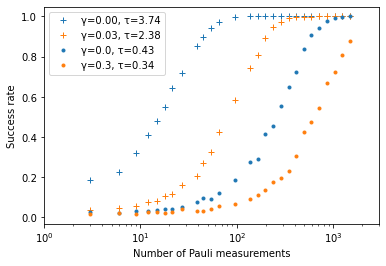}
    \caption{$W_1$ measurement state distinction success rates for a $99\%$ confidence threshold, with respect to the number of bipartite Pauli measurements, for noiseless and noisy scenarios. The crosses correspond to the original separation distance $d = 450 \mum$ while the dots correspond to $d = 350 \mum$.}
    \label{fig:success_PRL}
\end{figure}

To determine what value of $\lambda_a$ is sufficiently high, what we would aim to achieve in the SWP is to minimize false positives, that is, to have a small significance level $\alpha$ or equivalently a high confidence level $1-\alpha$. For a desired significance level $\alpha$ we define the minimum $\lambda_{\text{min}}$ by
\begin{equation}
    \mathbb{P}(\lambda_0 \geq \lambda_{\text{min}}) = \alpha,
\end{equation}
where $\lambda_0$ is the likelihood ratio assuming the null hypothesis.
In practice, we generate multiple data vectors $\underline{n}_0$ assuming the null hypothesis is true, and use the distribution of the resulting $\lambda_0$ for different values of $N$. For $\alpha = 1\%$, $\lambda_{\text{min}}$ is then the $99$-th percentile of the obtained $\lambda_0$. Once the $\lambda_{\text{min}}(N, \gamma, \tau(\gamma))$ are determined, one can generate the $\lambda_a$ from data where the alternative hypothesis is assumed true and inspect the frequency of $\lambda_a \geq \lambda_{\text{min}}$. This frequency is termed state distinction success rate, and is what has been plotted in Fig.~\ref{fig:success_PRL} for a confidence level of $99\%$, for the original parameter settings, and for a closer separation $d = 350 \ \mum$. For the original separation $d = 450 \ \mum$, in the noiseless case as well as with decoherence rate $\gamma = 0.03 \ \text{s}^{-1}$, $10^2$ measurements of $W_1$, obtained with $3\times 10^2$ repetitions of the experiment is enough to consistently rule out $\rho_0$. In the closer separation setting $d = 350 \ \mum$ certifying the alternative state reliably, requires around $10^3$ repetitions.

\begin{figure}
    \centering
    \includegraphics[scale=0.55]{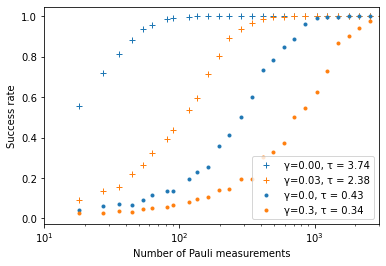}
    \caption{Tomographic state distinction success rates, for a $99\%$ confidence threshold, with respect to the number of bipartite Pauli measurements, for noiseless and noisy scenarios. The crosses correspond to the original separation distance $d = 450 \ \mum$ while the dots correspond to $d = 350 \ \mum$.}
    \label{fig:tomographies}
\end{figure}

\begin{figure}
    \centering
    \includegraphics[scale=0.55]{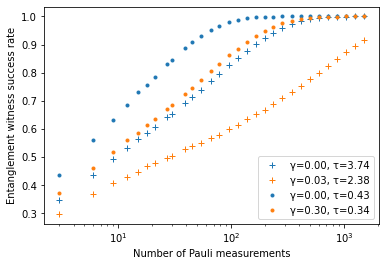}
    \caption{Probability of observing a negative empirical $W_1$ witness average, with respect to the number of bipartite Pauli measurements, for noiseless and noisy scenarios. The crosses correspond to the original separation distance $d = 450 \mum$ while the dots correspond to $d = 350 \mum$.}
    \label{fig:witneg}
\end{figure}

As stated previously, one must also be able to certify entanglement from the witness empirical data. We see from Fig.~\ref{fig:witneg} that even if, with $d = 450 \ \mum$, $10^2$ and $\gamma = 0.03 \ \text{s}^{-1}$,  $10^2$ witness measurements is sufficient to rule out $H_0$, there is only a $70\%$ chance for entanglement to be certified. Conversely, in the closer separation setup $d = 350 \ \mum$ the resulting state is more entangled, which makes entanglement certification more likely to succeed, but ruling out $H_0$ is more demanding.
In any case, this shows that the protocol does not need to be limited to the negligible CP coupling regime.

Hence, from the repeated measurement of a single entanglement witness, provided good enough knowledge of non-gravitational interactions, one can confirm the presence of an entangled state that could not have been obtained without gravity.

One could ask whether quantum state tomography would be more reliable for state distinction. In fact, to rule out $H_0$, the witness measurement and the full tomography are equivalently efficient even in a regime dominated by CP interactions, that is, with the separation $d$ brought down to $350\ \mum$, as shown by the dotted plots in Fig.~\ref{fig:success_PRL} and Fig.~\ref{fig:tomographies}. In both tomographic and witness measurement cases, the CP limit can be overcome, and distinguishing the two states reliably requires around $10^3$ bipartite Pauli measurements, that is, around $3\times 10^2$ witness measurements, or $10^2$ state tomographies. It seems the witness measurement works well enough not to need tomography. We shall discuss in Sec.~\ref{sec:reconstruction} the potential advantage to using tomographic data.

\section{Uncertainty in non-gravitational interactions}\label{sec:unknownforces}

In order to rule out $H_0$, we have used a likelihood ratio approach, and doing so we have assumed good knowledge of the non-gravitational interactions. Here, we show that the protocol with the likelihood ratio method is robust even when there is uncertainty in the non-gravitational coupling constants, with no need for further manipulation such as performing the experiment with two different separation distances.

The CP coupling constant $\alpha(R,\epsilon)$ is a good example of a quantity that is not precisely known, as some uncertainty may simply arise from the geometry of the not strictly spherical microdiamonds. At first glance from Eq.~\eqref{eq:altSDM} it seems that an uncertainty in the CP interaction could potentially account for the observed data assuming $H_a$, ruining all hopes of ruling out modified but plausible versions of $H_0$, in which $\alpha$ is modified from its predicted value. The potential issue is that the witness expectation values $\ev{W_1}_0$ and $\ev{W_1}_a$ measured on the two possible states after a fixed free-fall duration $\tau$ may coincide if $\alpha$ is modified.

Although for the original separation distance $d = 450 \ \mum$, this would require a relative uncertainty on $\alpha$ of over $500\%$, not having precise knowledge of $\alpha$ becomes more problematic at the proposed smaller separation setting $d = 350 \ \mum$. With this closer separation setting and a decoherence rate of $\gamma = 0.3 \ \text{s}^{-1}$ the proposed optimal free-fall duration for witnessing entanglement is $\tau = 0.34 \ \text{s}$. Numerics show that for this same free-fall duration, a modified null hypothesis state where $\alpha' = 1.087\alpha$ will yield the same witness expectation value. This means a less than $9\%$ uncertainty on $\alpha$ is sufficient for witness measurement values in the null-hypothesis state to converge to the same value as in the alternative hypothesis. One known solution to solve this issue, sometimes referred to as a differential measurement, is to use the fact that the CP potential follows a $1/r^7$ law, whereas the gravitational follows $1/r$, which should show if we repeat the experiment but slightly increase the separation distance.

However, we do not need to change the separation at all.
\begin{figure}
    \centering
    \includegraphics[scale=0.55]{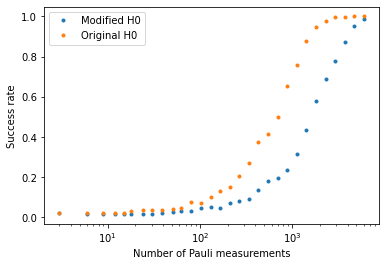}
    \caption{Witness measurement success rates with $d = 350 \ \mum, \gamma = 0.3 \ \text{s}^{-1}$  of ruling out $H_0$ or the modified $H_0$ in favour of $H_a$, with respect to the number of bipartite Pauli measurements.}
    \label{fig:differential}
\end{figure}
Interestingly, the protocol is already itself a differential measurement, and although one can think of a modified $\alpha$ that could make the null and alternative witness expectation values coincide, discrepancies will still show in the individual Pauli observables involved in the witness measurement. This is corroborated by the plot shown in Fig.~\ref{fig:differential} where we observe that the success rate still increases with the number of measurements, despite having performed $\alpha \to 1.087\alpha$ for the null hypothesis state, such that the witness expectation values are equal in both hypotheses. This modification of $\alpha$ in the null hypothesis does makes it more demanding to rule out $H_0$ but not impossible. The number of bipartite Pauli measurements for a good state distinction success rate goes from a few $10^3$ to a little less than $10^4$. 

In fact, the spin density matrices in the null and alternative hypotheses can never be made equal by any change in the coupling constants. This is because in Eq.~\eqref{eq:altSDM} the $Q^{(1)}$ and $Q^{(7)}$ quantities are non-proportional tensors, they are differences between several proximities to different powers. Let us note that the same observation applies if one wishes to include any potential that is not $1/r$, such as a dipole-dipole interactions, which would involve a $Q^{(3)}$ quantity that is linearly independent of $Q^{(1)}$ and $Q^{(7)}$. This emphasises the power of our approach based on Pauli measurement likelihood ratios, rather than comparisons of only the witness expectation values. In the latter case one cannot distinguish these different interactions, whereas the former can.

\section{State reconstruction}\label{sec:reconstruction}
Finally, we address a potential loophole for the witness-based approach and describe how full tomography can provide a solution. We present results from tomographic simulations which provide an order of magnitude for the number of measurements required.

We have shown that by analyzing the witness Pauli measurements, we are able to consistently state with a high degree of statistical confidence not only that the state is entangled but also that the state was produced by a gravitational interaction as opposed to merely a CP interaction. This would be a highly significant observation, but there remains a potential loophole. Since entanglement witnesses are not entanglement monotones, a skeptic could argue that we have not explicitly shown that gravity has \emph{increased} the entanglement. Indeed, the null hypothesis state is already entangled, with negativity $\mathcal{N}(\rho_0) > 0$, and there exist other valid quantum states $\rho_a'$ that are indistinguishable from $\rho_a$ in the witness statistics, the negativities of which may satisfy $\mathcal{N}(\rho_0) > \mathcal{N}(\rho_a') > 0$. One such loophole state is presented explicitly in appendix~\ref{sec:proofs3}. However contrived the argument would have to be to justify such a state, an answer to eliminate this loophole can be provided by using full tomography to calculate an entanglement monotone, as we shall demonstrate.

Quantum state tomography is a method to estimate quantum states from a complete set of measurements on many copies of the same state. It has been widely studied in general~\cite{nielsen2002quantum, christandl2012reliable} as well as in its application to entanglement verification~\cite{arrazola2013reliable}.
Tomographic data allows one to perform state reconstruction from which statistical statements regarding entanglement monotones can be made. We construct state estimators using the well known method of maximum likelihood estimation~\cite{banaszek1999maximum}. It may not always be the most accurate estimation method~\cite{ferrie2018maximum}, but it is sufficient for our purpose, and has been extensively used in experiments~\cite{roos2004bell, blatt2008entangled, lvovsky2009continuous}.

We seek to predict how reliably the experiment with tomographic data can certify gravitationally induced entanglement growth. To this end, we simulate a series of full tomographies on $\rho_0$ and $\rho_a$, and reconstruct their corresponding maximum likelihood states $\rho_{\text{ML}}$ following a fixed-point iterative method~\cite{hradil1997quantum}. Explicitly, the maximum likelihood state can be obtained from the empirical data vector $\underline{n} = [n_i]$ containing the number of occurrences of the Pauli-measurement outcome $\dyad{e_i}{e_i}$ as it solves $\rho_{\text{ML}} = \hat{R}(\rho_{\text{ML}})\rho_{\text{ML}}$, where
\begin{equation}
    \hat{R} : \rho \longmapsto \frac{1}{||\underline{n}||_1}\sum_i \frac{n_i}{\Tr(\rho\dyad{e_i}{e_i})}\dyad{e_i}{e_i}.
\end{equation}
Then the sequence of two-qubit density matrices defined by $\rho_0  =  I_4/4$ and $\forall k\in\mathbb{N}, \rho_{k+1} =\mathcal{N}_{\Tr}\left(\hat{R}(\rho_k)\rho_k\hat{R}(\rho_k)\right)$
where $\mathcal{N}_{\Tr}$ designates trace normalization, converges heuristically to the maximum-likelihood state. In our simulations, we end the algorithm at the $100$th iteration. Among $10^3$ full tomographic state reconstructions using $9\times 10$ measurements, the fidelity~\cite{jozsa1994fidelity} $\mathcal{F}(\rho_{100}, \rho) = \left(\Tr(\sqrt{\sqrt{\rho_{100}}\rho\sqrt{\rho_{100}}})\right)^2$ between the simulated states is at least $90\%$ and on average $95\%$. Among $10^3$ full tomographic state reconstructions using $9\times 10^3$ measurements, the fidelity is always over $99\%$.

From the maximum likelihood reconstructed states, we find different negativity distributions in the two hypotheses. The results are summed up in Fig.~\ref{fig:tomographysuccess}, and show that in the original separation $d = 450 \ \mum$ a few $10^3$ Pauli measurements is enough to consistently reject the null hypothesis and loophole states when $\gamma = 0.03 \ \text{s}^{-1}$, and for the closer setup $d = 350 \ \mum$ where the CP interaction becomes significant, $10^4$ Pauli measurements is sufficient to reject the null hypothesis and loophole sates around $90\%$ of the time when $\gamma = 0.3 \ \text{s}^{-1}$.

\begin{figure}
    \begin{tabular}{|c| c | c| c| c| c|} \hline
        \multicolumn{2}{|c|}{\diagbox{Setting}{Repetitions}} & $9$ & $90$ & $9\times 10^2$ & $9\times 10^3$ \\ \hline
        \multirow{2}{*}{$ d=450 \mum$} & $\gamma = 0$ & $1.8\%$ & $40.7\%$ & $>99.9\%$ & $>99.9\%$ \\ \cline{2-6}
        & $\gamma = 0.03$ & $2.6\%$ & $2.8\%$ & $66.3\%$ & $>99.9\%$ \\ \hline
        \multirow{2}{*}{$ d=350 \mum$} & $\gamma = 0$ & $1.2\%$ & $7.5\%$ & $53.2\%$ & $>99.9\%$ \\ \cline{2-6}
         & $\gamma = 0.3$ & $1.7\%$ & $4.1\%$ & $10.2\%$ & $88.7\%$ \\ \hline
    \end{tabular}
    \caption{Probability for the reconstructed alternative hypothesis state to have a higher negativity than the $99$-th percentile most entangled reconstructed null hypothesis state, with respect to the number of Pauli measurements. Results are shown for the original and shorter separation distances, and for the noiseless and strongest decoherence rate cases.}
    \label{fig:tomographysuccess}
\end{figure}
It should therefore be possible, by reconstructing the state from tomographic data on a relatively large but not unreasonable number of measurements, to obtain reliable proof of entanglement growth by gravitational interaction even in the presence of other stronger and possibly ill-known couplings, and decoherence.

\section{Conclusion}
Recent proposals such as the SWP~\cite{bose2017spin} for a test of the non-classical nature of gravity feature a promising protocol. Although the experimental requirements are futuristic, some important theoretical questions, such as the plausibility of using position eigenstates, the possibility of a better witness, the necessity of a CP closest approach limit, and the significance of experimental data, were left open. In this paper, we have addressed those questions and shown that the position eigenstate is a valid approximation, for the calculation of entanglement, to a good range of thermal states in the predicted experiment durations. By introducing an optimal entanglement witness, we were able to show that entanglement revelation is possible in less free-fall time, with higher decoherence rates. We have furthermore demonstrated that a statistical approach to the witness method decomposed into Pauli measurements allowed to discern gravitational from uncertain non-gravitational contributions. Finally, we have spelled out and closed a potential loophole in the witness based approach to establishing the entangling capacity of the gravitational interaction.

In addition to loosening the regime and duration constraints of the SWP, the methods laid out in this work may benefit to further discussions around similar tests of quantum physics, such as precision measurements of the CP effect, and probing the fifth force.

\vspace{5mm}
\begin{acknowledgments}
We acknowledge insightful discussions with Benjamin Stickler, Hyukjoon Kwon, Sougato Bose and Anupam Mazumdar. This work was funded by the ESPRC Centre for Doctoral Training in Controlled Quantum Dynamics, and QuantERA ERA-NET Cofund in Quantum Technologies implemented within the European Union's Horizon 2020 Programme and KIAS visiting professorship. MSK acknowledges the Royal Society.
\end{acknowledgments}

\bibliographystyle{apsrev4-1}
\bibliography{main}

\newpage
\onecolumngrid
\begin{appendix}

\section{Form of the spin density matrix}\label{sec:proofs1}
We give an explicit derivation of Eq.~\eqref{eq:thermal1} and Eq.~\eqref{eq:thermal2}. Given two copies of an arbitrary initial state for which we choose a $P$-function representation
\begin{equation}
    (\pi\otimes\pi) = \iint d^2\varepsilon d^2\zeta P(\varepsilon)P(\zeta)\dyad{\varepsilon\zeta}{\varepsilon\zeta},
\end{equation}
the spin density matrix is deduced from Eq.~\eqref{eq:SDMelements} to read
\begin{equation}\begin{aligned}
    s_{\alpha\beta\mu\nu} = \frac{1}{4} e^{-iGm^2\tau Q^{(1)}_{\alpha\beta\mu\nu}/\hbar}\iint d^2\varepsilon d^2\zeta P(\varepsilon)P(\zeta)\Tr[{\hat{\slashed{U}}_{\alpha\beta}}^{\dagger}\hat{\slashed{U}}_{\mu\nu}\dyad{\varepsilon\zeta}{\varepsilon\zeta}].
    \end{aligned}
\end{equation}
From \eqref{eq:propproduct} it follows that
\begin{equation}\begin{aligned}
    s_{\alpha\beta\mu\nu} = \frac{1}{4} e^{-iGm^2\tau Q^{(1)}_{\alpha\beta\mu\nu}/\hbar}e^{-iG^2m^3\tau^3Q^{(4)}_{\alpha\beta\mu\nu}/6\hbar}\iint d^2\varepsilon d^2\zeta P(\varepsilon)P(\zeta)\braket{\varepsilon}{\varepsilon+\theta_{\alpha\beta\mu\nu}}\braket{\zeta}{\zeta - \theta_{\alpha\beta\mu\nu}}.
    \end{aligned}
\end{equation}
The remaining double integral can be simplified as

\begin{equation}
\begin{aligned}
     \iint d^2\varepsilon d^2\zeta P(\varepsilon)P(\zeta)\braket{\varepsilon}{\varepsilon+\theta}\braket{\zeta}{\zeta - \theta} &= e^{-|\theta|^2}\left(\int d^2\varepsilon P(\varepsilon)e^{\frac{1}{2}(\varepsilon^*\theta - \theta^*\varepsilon)}\right) \left(\int d^2\zeta P(\zeta)e^{\frac{1}{2}(\theta^*\zeta - \zeta^*\theta)}\right),\\
    &= e^{-|\theta|^2}\tilde{P}\left(\frac{\theta}{2}\right)\tilde{P}\left(-\frac{\theta}{2}\right),\\
    &= e^{-|\theta|^2}\left(C_N\left(\frac{\theta}{2}\right)\right)^2,
\end{aligned}
\end{equation}
where we have dropped the indices for $\theta$ and used the fact that the $P$-function is real and so has even Fourier transform $\tilde{P}$, and where $C_N$ is the normally ordered characteristic function of the initial local state $\pi$. Explicitly $C_N(\lambda) = \Tr[\pi e^{\lambda\crea{a}}e^{-\lambda^*\hat{a}}]$. This gives Eq.~\eqref{eq:thermal1}.

If the initial local state is thermal with $\langle\hat{N}\rangle = \overline{n}$, then the characteristic function is $C_N(\lambda) = e^{-\overline{n}|\lambda|^2}.$ This can be straightforwardly shown, starting from $P(\varepsilon)=\frac{e^{-|\varepsilon|^2/\overline{n}}}{\pi\overline{n}},$ and evaluating the integrals as we now demonstrate for completeness.

By writing $\varepsilon=x+iy$ and $\theta = a+ib$ we have
\begin{equation}
\begin{aligned}
    C_N\left(\frac{\theta}{2}\right)=\frac{1}{\pi\overline{n}}\int d^2\varepsilon e^{-|\varepsilon|^2/\overline{n} + \frac{1}{2}(\varepsilon^*\theta - \theta^*\varepsilon)} &= \frac{1}{\pi\overline{n}}\int dx e^{-x^2/\overline{n}+ ixb} \int dy e^{-y^2/\overline{n} + iya},\\
    &= \frac{1}{\pi\overline{n}}\int dx e^{-(x-i\overline{n}b/2)^2/\overline{n}}e^{-\overline{n}b^2/4} \int dy e^{-(y-i\overline{n}a/2)^2/\overline{n}}e^{-\overline{n}a^2/4},\\
    &= e^{-\overline{n}|\theta|^2/4}.
\end{aligned}
\end{equation}
For the second line we have completed the square in the exponents and for the third we use the result for the standard Gaussian integral. Thus we establish Eq.~\eqref{eq:thermal2}.

\section{Decoherence rate limit}\label{sec:proofs2}
Recall
\begin{equation}
    W_1 = I\otimes I - X\otimes X - Z\otimes Y - Y\otimes Z.
\end{equation}
Then for a Hermitian 4 by 4 matrix $\rho = (\rho_{ij})_{1\leq i,j \leq 4}$ one can compute
\begin{equation}
    \Tr(W_1\rho) = \Tr(\rho) + 2\Im{\rho_{12}} + 2\Im{\rho_{13}} - 2\Re{\rho_{14}} - 2\Re{\rho_{23}} - 2\Im{\rho_{24}} - 2\Im{\rho_{34}}.
\end{equation}
With decoherence rate $\gamma$, the spin state after free fall of duration $t$ reads
\begin{equation}
    \rho = \frac{1}{4}\begin{pmatrix} 1 & e^{-\gamma t}e^{-i\Delta\phi_{LR}} & e^{-\gamma t}e^{-i\Delta\phi_{RL}} & e^{-2\gamma t} \\ e^{-\gamma t}e^{i\Delta\phi_{LR}} & 1 & e^{-2\gamma t}e^{i(\Delta\phi_{LR} - \Delta\phi_{RL})} & e^{-\gamma t}e^{i\Delta\phi_{LR}} \\ e^{-\gamma t}e^{i\Delta\phi_{RL}} & e^{-2\gamma t}e^{i(\Delta\phi_{RL} - \Delta\phi_{LR})} & 1 & e^{-\gamma t}e^{i\Delta\phi_{RL}} \\ e^{-2\gamma t} & e^{-\gamma t}e^{-i\Delta\phi_{LR}} & e^{-\gamma t}e^{-i\Delta\phi_{RL}} & 1
    \end{pmatrix}
\end{equation}
Then the witness expectation value is given by
\begin{equation}
    \Tr(W_1\rho) = 1 - e^{-\gamma t}\left(\sin(\Delta\phi_{LR}) + \sin(\Delta\phi_{RL})\right) - \frac{1}{2} e^{-2\gamma t}(1 + \cos(\Delta\phi_{LR} - \Delta\phi_{RL})).
\end{equation}
Writing $\omega_{\mu\nu}t = \Delta\phi_{\mu\nu}$ a first order expansion around $t = 0$ gives
\begin{equation}
    \Tr(W_1\rho) = (2\gamma - (\omega_{LR} + \omega_{RL}))t + o(t),
\end{equation}
so there is no more immediately witnessed entanglement for decoherence rates greater than the average of the two path coupling frequencies $\gamma \geq (\omega_{RL} + \omega_{LR})/2$.

\section{Witness loophole state}\label{sec:proofs3}
We simulated a $3\times 10^3$ element string of Pauli outcome data obtained from $10^3$ measurements of $W_1$ on $\rho_a$. Given this data, we have constructed a state $\rho_a'$ that is at least as likely as $\rho_a$ given the data but has a negativity that is lower than that of the null hypothesis state $\rho_0$. Explicitly for separation distance $d = 350 \ \mum$, decoherence rate $\gamma = 0.3 \ \text{s}^{-1}$ with corresponding free-fall duration $\tau = 0.34 \ \text{s}$, one has $\mathcal{N}(\rho_0) \approx 0.108 > \mathcal{N}(\rho_a') \approx 0.104$ where with a rounding to $3$ significant figures
\begin{equation}\begin{aligned}
    \rho_a' =
    \begin{pmatrix} 0.256 & 0.009+ 0.012i &   0.042-0.174i &
   0.212+0.010i \\
  0.009-0.012i &  0.244 &         0.109-0.022i & 
  -0.008+0.004i \\
 0.042+0.174i &  0.109+0.023i &  0.246 &
   0.017+0.161i \\
  0.212-0.011i & -0.008-0.004i &  0.017-0.161i &
   0.254
   \end{pmatrix}.
   \end{aligned}
\end{equation}
This state was obtained via a constrained optimization method (sequential least squares programming) over all valid $2$-qubit quantum states. The function to minimize was the negative logarithmic likelihood ratio given the data, under the constraint that the state should be less negative than the null hypothesis state.

For completeness we present the $2$-qubit state space parametrization. The parametrization was arrived at by using the Cholesky decomposition of positive-semidefinite density matrices as $\rho=LL^{\dagger},$ where $L$ is a lower diagonal matrix with real diagonal coefficients which we write as
\begin{equation*}
    L=\begin{pmatrix}
    l_1 & 0 & 0 & 0\\
    l_5 & l_2 & 0 & 0\\
    l_8 & l_6 & l_3 & 0\\
    l_{10} & l_9 & l_7 & l_4
    \end{pmatrix}.
\end{equation*}
This defines $16$ real parameters, $4$ for the diagonal terms and $12$ for the off-diagonals. The unit trace condition $\text{Tr}(\rho)=\text{Tr}(LL^{\dagger})$ gives $\sum_{i=1}^{10} |l_{i}|^2=1.$ This defines the surface of an $n$-sphere (specifically a $9$-sphere), so we parametrize with $l_1=\cos\theta_1,$ $l_2=\sin\theta_1\cos\theta_2,$ ... $l_{9}=(e^{i\phi_{9}}\prod_{i=1}^{8}\sin\theta_i )\cos\theta_{9},$ $l_{10}=e^{i\phi_{10}}\prod_{i=1}^{9}\sin\theta_i.$ Note $l_1,...l_4$ do not require $e^{i\phi}$ phase terms since they are real. This therefore parametrizes the full $2$-qubit state space with $15$ angles, $\theta_1,...\theta_{9},$ and $\phi_{5},...,\phi_{10}.$

\end{appendix}

\end{document}